\documentclass[aip,reprint,amsmath,amssymb]{revtex4-2}

\usepackage{color}
\usepackage{graphicx}
\usepackage{amsmath}
\usepackage{subcaption}
\usepackage{amssymb}
\usepackage{lineno}
\usepackage{natbib}

\newcommand{\lei}{\lambda_{ei}}
\newcommand{\vth}{v_{th}}

\graphicspath{{figs/}}

\usepackage{placeins}

\begin{document}
	\title{Plasma transport by carrier electrons: delocalization length for heat-flux}
	
	\author{C. A. Walsh}
	\email{walsh34@llnl.gov}
	\affiliation{Lawrence Livermore National Laboratory, 7000 East Avenue, Livermore, CA 94550}

	\date{\today}
	

	\begin{abstract}
		The kinetic understanding of the three magnetized heat-flux components is unified in the linear regime with a full electron-electron collision operator and a converged harmonic expansion of the electron distribution (up to $f_6$). The delocalization length  $\lambda_d=\lambda_{ei} Z/\sqrt{Z+4.45}$ is an improved heat-flux non-locality metric accounting for the localizing effect of electron-electron collisions on the electron population carrying heat-flux. Magnetization also localizes the carrier electrons; the delocalization Hall parameter $\psi_d=13.4 \chi Z/(Z+4.45)$ shows that this occurs at a faster rate than magnetization of thermal conductivity ($\psi_{\kappa}= 9.94\chi(Z+0.17)/(Z+5.88)$). The traditional Knudsen number, which uses the electron-ion mean-free path, does not fairly represent non-locality across ionization and magnetization; $Kn_d = 8.28\lambda_{d\bot}/L_T$ should instead be used. Righi-Leduc suppression is found to be the square of the perpendicular suppression, with no additional length-scale required. This approach lays the groundwork for consistent treatment of kinetic plasma transport, with subsequent papers addressing the kinetically-modified induction equation. 
	\end{abstract}
	\maketitle
	
	Classical theory describes fluxes of electron energy in a 'local' plasma through the transport coefficients \cite{spitzerTransportPhenomenaCompletely1953,braginskiiTransportProcessesPlasma1965,epperleinPlasmaTransportCoefficients1986,sadlerSymmetricSetTransport2021,daviesTransportCoefficientsMagneticfield2021,walsh_paper0}. A temperature gradient is assumed to not affect the isotropic $f_0$ distribution of electrons, only driving an anisotropic component $\mathbf{f}_1$ responsible for the heat-flux. As temperature gradients become steeper, $f_0$ is also perturbed. The metric typically used to assess this effect is the Knudsen number:
	\begin{equation}
		Kn = \frac{\lei}{L_T} \label{eq:Knudsen}
	\end{equation}
	where $L_T = T_e/|\nabla T_e|$ is the temperature length-scale and $\lei$ is the mean free path between electron-ion (e-i) collisions.
	
	If an electron travels further between collisions than the temperature length-scale ($\lei > L_T$), it is 'non-local'. As the high energy electrons have longer mean free paths (scaling as the 4th power of the electron velocity $v^4$), they leave the region of interest. The high-energy tail of the electron distribution, which contributes most to heat-flux, is depleted. Therefore, the heat-flux is lower than classical predictions; a flux-limiter is used in hydrodynamics codes as an attempt to account for this reduction\cite{malone1975}.
	
	The Knudsen number conflates two distinct properties of a temperature structure. For a sinusoidal perturbation of relative amplitude $\delta$ and wavenumber $k$, the steepest gradient gives $\lambda_{ei}/L_T = \delta\,k\lambda_{ei}$: large $Kn$ can arise from a short wavelength at small amplitude, or from a long wavelength at large amplitude. The former is \emph{non-locality}, measured by $k\lambda$: whether the flux at a point depends on the gradient there or on its surroundings. The latter is \emph{non-linearity}, measured by $\delta$: how much the plasma itself (e.g. mean free path) varies across the structure. Classical transport is non-linear but exactly local; kinetic effects require non-locality. This paper isolates the non-locality axis, using small-amplitude perturbations for which the kinetic response is linear and can be expressed as a suppression factor per Fourier mode.
	
	The first question this paper seeks to address is whether the e-i mean free path in equation \ref{eq:Knudsen} is the appropriate length-scale to assess non-locality; for low-$Z$ plasmas this length should be lowered by electron-electron (e-e) collisions\cite{luciani1983,epperleinPracticalNonlocalModel1991,brodrickIncorporatingKineticEffects2018,daviesNonlocalSuppressionBiermann2023,walshKineticCorrectionsHeatflow2024,lau2025}. Using $Kn$ as an input to flux-limiters will therefore not consistently suppress different plasma species. For this purpose, a delocalization length $\lambda_{d}$ is measured from kinetic simulations, as well as the rate of suppression.
	
	Then there is the question of magnetic fields. While some experiments purposefully apply these fields\cite{moodyIncreasedIonTemperature2022,changFusionYieldEnhancement2011}, they are also spontaneously generated and are increasingly considered ubiquitous to plasma physics experiments\cite{liMeasuringFieldsLaserProduced2006,willingaleFastAdvectionMagnetic2010,farmerSimulationSelfgeneratedMagnetic2017,walshSelfGeneratedMagneticFields2017}. Magnetization results in thermal conduction becoming anisotropic:
	\begin{equation}
		\mathbf{q}_e = -\kappa_{\parallel} \nabla_{\parallel}T_e -\kappa_{\bot} \nabla_{\bot}T_e - \kappa_{\wedge} \hat{\mathbf{b}} \times \nabla T_e  \label{eq:magheatflow}
	\end{equation}
	with unrestricted heat-flow along magnetic field lines ($\kappa_\parallel$) and restricted heat-flow perpendicular ($\kappa_\bot$) as the Hall parameter $\chi=\omega_e \tau_{B}$ (a measure of magnetization) increases\cite{braginskiiTransportProcessesPlasma1965,epperleinPlasmaTransportCoefficients1986}. The $\kappa_{\wedge}$ term is Righi-Leduc heat-flow, which can be conceptualized as thermal conduction being rotated by the magnetic field. Note that $\tau_B=3\sqrt{\pi}\tau_{ei}/4$ is the electron collision time defined by Braginskii\cite{braginskiiTransportProcessesPlasma1965}.

	\begin{figure}[tb]
		\includegraphics[width=\linewidth]{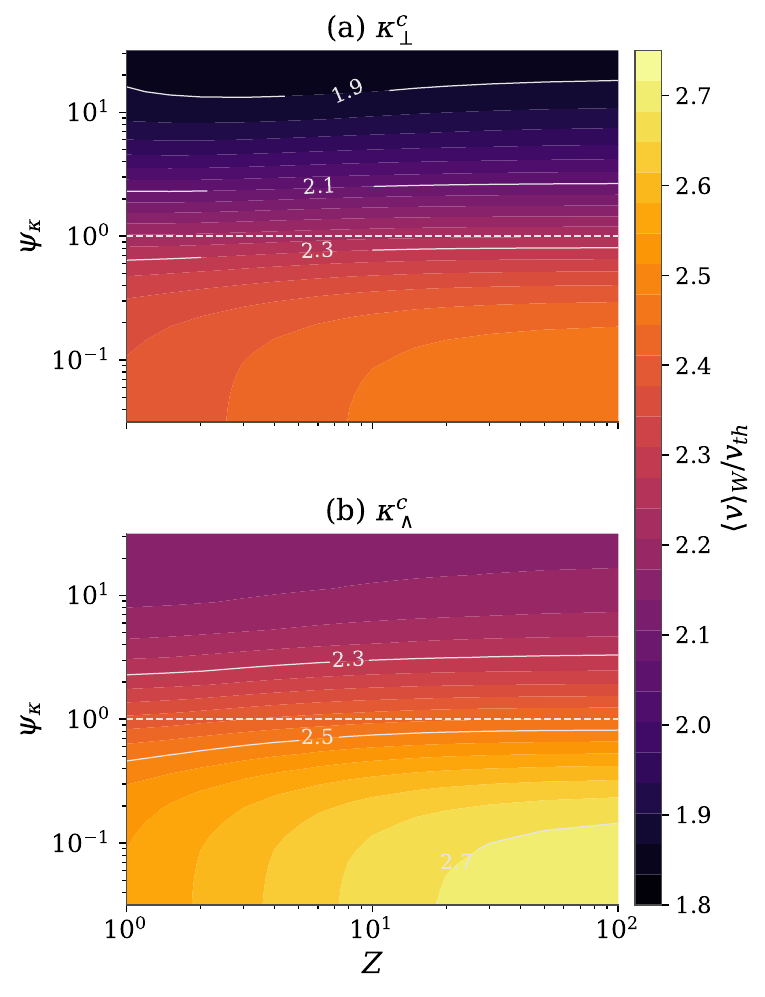}
		\caption{\label{fig:carrier_map} Average electron velocity that carries perpendicular heat-flow (a) and Righi-Leduc (b) for a classical plasma. The y-axis is the carrier Hall parameter from equation \ref{eq:psi_kap}. The x-axis is plasma ionization. Both magnetization and e-e collisions lower the energy of electrons responsible for heat-flux.}
	\end{figure}

	How is the delocalization length modified by magnetization? Are all components suppressed the same? And how does this affect how magnetohydrodynamic (MHD) codes should flux-limit the different components of magnetized heat-flow?
	
	The answer to these questions is sought in the regime of 'marginal non-locality', where corrections to classical transport begin to emerge. As will be seen, it is possible to thoroughly characterize the linear response of this regime; future work extends this approach to the magnetic induction equation.
	
	The approach taken is through the lens of carrier electrons: studying the electron population responsible for carrying heat-flux. This concept was developed to explain the behavior of the classical transport coefficients\cite{walsh_paper0} (the previous paper in the series). A review of the important findings applicable to the transition into non-locality are included in section \ref{sec:carrier}.
	
	Section \ref{sec:theory} walks through transport theory by outlining a hierarchy of length-scales responsible for classical and kinetic plasmas. Section \ref{sec:VFP} presents the Vlasov Fokker-Planck (VFP) simulation methodology, including an introduction to VFP theory. The simulations are considered 'state of the art' (harmonic expansion up to $f_6$, full e-e collision operator\cite{tzoufrasVlasovFokkerPlanck2011}), and are applied to the most fundamental kinetic problem: 1D linear temperature perturbations. 
	
	Sections \ref{sec:unmag} and \ref{sec:mag} first demonstrate how e-e collisions and magnetization affect the susceptibility of a plasma to kinetic effects. They then fit the unmagnetized and magnetized delocalization lengths to a database of $>10^5$ linear-mode calculations, which were validated against full VFP simulations; this database spans ionization, magnetization and degree of non-locality. The results are summarized in table \ref{tab:lengths}. The suppression factors for parallel, perpendicular and Righi-Leduc heat-flux are shown to compare favorably with the full VFP simulations.  Section \ref{sec:FL} then applies these new lengths to a magnetized heat-flow flux limiter, for implementation into MHD codes.

	Appendix \ref{sec:regime} discusses the regime of applicability of these results. Convergence of the harmonic expansion up to $f_6$ is demonstrated in appendix \ref{sec:regime_trunc}. The superposition of multiple linear modes are shown to hold in appendix \ref{sec:regime_superposition}. Appendix \ref{sec:eigensolver} describes the linear mode solver. Appendix \ref{sec:kinetic_carrier} extends the fit for the heat-flow classical carrier velocity\cite{walsh_paper0} into the kinetic regime.

	\begin{table}[tb]
		\caption{Kinetic suppression of the three magnetized heat-flux components. All components share one delocalization length $\lambda_d$ and one delocalization Hall parameter $\psi_d$. Righi-Leduc is the square of the perpendicular suppression function. Classical coefficients are from the previous paper in the series, with $\psi_{\kappa}$ being the carrier Hall parameter for heat-flux\cite{walsh_paper0}.}
		\label{tab:lengths}
		\begin{ruledtabular}
			\renewcommand{\arraystretch}{1.7}
			\setlength{\tabcolsep}{10pt}
			\begin{tabular}{llll}
				& Classical & \shortstack[l]{Non-locality\\metric} & \shortstack[l]{Kinetic\\suppression} \\
				\colrule
				$\kappa_\parallel$ & $\kappa_\parallel^c = 13.58\frac{Z+0.26}{Z+4.31}$ & $k\lambda_d$ & $\Gamma_q(k\lambda_d)$ \\
				$\kappa_\perp$ & $\kappa_\parallel^c/(1+\psi_\kappa^{1.5})$ & $k\lambda_{d\perp}$ & $\Gamma_q(k\lambda_{d\perp})$ \\
				$\kappa_\wedge$ & peaks at $\psi_\kappa\approx0.92$ & $k\lambda_{d\perp}$ & $\Gamma_q(k\lambda_{d\perp})^2$ \\
				\colrule
				\multicolumn{4}{l}{\rule{0pt}{6ex}$\lambda_d=\lambda_{ei}\frac{Z}{\sqrt{Z+4.45}}$, \qquad
					$\lambda_{d\bot}=\frac{\lambda_d}{1+\psi_d^{1.6}}$, \qquad
					$\psi_d=13.4\,\chi\frac{Z}{Z+4.45}$} \\[6pt]
				\multicolumn{4}{l}{$\psi_\kappa=9.94\,\chi\,\frac{Z+0.17}{Z+5.88}$, \qquad
					$\Gamma_q(x)=\dfrac{1+27.29x}{1+27.29x+294.5x^2}$} \\
			\end{tabular}
		\end{ruledtabular}
	\end{table}

	\section{Carrier electrons for classical transport}
	\label{sec:carrier}
	
	A 'typical' electron in a classical plasma travels at the thermal velocity $v_{th}=\sqrt{2T_e/m_e}$. However, these are not the electrons responsible for carrying the heat-flux. For an unmagnetized Lorentz plasma ($Z=\infty$) the carrier-averaged velocity for heat-flux is $2.48v_{th}$. e-e collisions (at finite $Z$) act to lower the thermal conductivity, as well as reduce the carrier velocity to $2.38v_{th}$ for $Z=1$.\cite{walsh_paper0}
	
	The carrier population also depends on magnetization. Magnetization can be quantified per electron velocity as:
	\begin{equation}
		\chi(v) = \frac{\omega_e}{\nu_{ei}(v)}
		= \frac{4}{3\sqrt{\pi}}\,\left(\frac{v}{v_{th}}\right)^3 \chi ,
		\label{eq:chi_v}
	\end{equation}
	where $\omega_e=e|\mathbf{B}|/m_e$ is the cyclotron frequency, $\nu_{ei}$ is the e-i collision frequency and $\chi$ is the bulk Hall parameter typically used to quantify magnetization\cite{braginskiiTransportProcessesPlasma1965,sadlerSymmetricSetTransport2021,daviesTransportCoefficientsMagneticfield2021}.
	
	$\chi=1$ corresponds to electrons at velocities of $v/v_{th} = 1.10$ or greater being magnetized. However, as the heat-flux carrier electrons are at velocities much greater than this, $\chi$ does not fairly represent thermal conductivity magnetization. In fact, due to e-e collisions, low $Z$ plasmas are harder to magnetize than high $Z$. For $Z=1$, $\kappa_{\bot}^c(\chi=1)/\kappa_{\parallel}^c = 0.31$; this falls to $0.048$ for $Z=\infty$.
	
	To this end, a carrier Hall parameter for thermal conductivity was defined\cite{walsh_paper0}:
	\begin{equation}
		\psi_{\kappa} = 9.94 \frac{Z+0.17}{Z+5.88} \omega_e \tau_B \label{eq:psi_kap}
	\end{equation}
	where $\psi_{\kappa}=1$ corresponds to 50\% thermal conductivity suppression, independent of plasma ionization $Z$.
	
	The preceding work took this approach for all the classical transport processes (e.g. thermal conduction, Nernst advection, or resistive diffusion of magnetic fields)\cite{walsh_paper0}. Each process is carried by a different carrier population, and therefore requires a different carrier Hall parameter $\psi$ to fairly represent magnetization.
	
	That work also calculated how the classical carrier velocities change with e-e collisions ($Z$) and magnetization ($\psi$). This dependence is plotted in figure \ref{fig:carrier_map} for (a) $\kappa_{\bot}^c$ and (b) $\kappa_{\wedge}^c$. Magnetization lowers the carrier velocity to $v/v_{th} = 1.85$ for $\kappa_{\bot}^c$ and $2.17$ for $\kappa_{\wedge}^c$, independent of ionization.

	What can we learn about kinetic heat-flux from this classical carrier approach? The high velocity electrons responsible for kinetic effects are less collisional than the bulk. Any process that lowers the carrier velocity (e-e collisions and magnetization) will therefore make a plasma more localized. The carrier velocity for Righi-Leduc ($\kappa_{\wedge}^c$) was also found to be greater than perpendicular heat-flow ($\kappa_{\bot}^c$); it is therefore expected to be more susceptible to kinetic suppression.

	\section{Kinetic transport as a hierarchy of length scales}
	\label{sec:theory}
	
	\subsection{Unmagnetized length scales}
	\label{sec:unmag_length}
	
	When an electron collides with an ion, its energy remains unchanged due to the large mass ratio ($>1800$). Instead, the electrons are elastically scattered such that they lose directionality. The electron-ion mean free path
	\begin{equation}
		\lambda_{ei}(v) \;=\; \frac{v}{\nu_{ei}(v)}
		\;=\; \frac{16 \pi \epsilon_0^2 T_e^2}{n_e Z e^4 \ln\Lambda} \left(\frac{v}{\vth}\right)^{\!4}
		\label{eq:lambdaei}
	\end{equation}

	can be conceptualized as the distance an electron travels before ion scattering makes it forget where it was originally going. After deflection it is still carrying its energy, only in a different direction. $v/v_{th}$ has been kept in this relation to emphasize how the faster electrons within the population have longer paths between scattering events than the slower electrons.
	
	Thermal conductivity is governed by how far an electron travels before losing its directionality \emph{due to both electrons and ions}, $\lambda_{\kappa}$. A Lorentzian ($Z=\infty$) plasma has an analytic classical thermal conductivity $\kappa_{\parallel}^c = 128/3\pi$\cite{spitzerTransportPhenomenaCompletely1953}, which is then lowered by e-e collisions \cite{walsh_paper0}.
	
	$\lambda_{\kappa}$ can be found by writing the unmagnetized heat-flow as $\mathbf{q} = -\lambda_{\kappa} n_e v_{th} \nabla T_e$\cite{walsh_paper0}:
	\begin{equation}
		\lambda_{\kappa} = 9.03 \lei\frac{Z+0.26}{Z+4.31} \label{eq:lam_kap}
	\end{equation}
	i.e. in a Lorentzian plasma the heat carriers travel a distance $9.03\lei$ between deflections, an order of magnitude further than a thermal electron due to the carrier population living at $2.48v_{th}$.
	
	A plasma is classical if an electron loses its energy within a distance $\lambda_d$ that is short compared with the temperature length-scale $L_T$. Only e-e collisions result in electrons losing their energy. The resultant path length is:
	\begin{equation}
		\lambda_{ee} = Z \lei
		\label{eq:lambdaee}
	\end{equation}
	Where the factor of $Z$ comes from the relative collision frequency. This is the distance an electron moves, not accounting for the fact that it is redirected numerous times in the process. The net displacement before an electron loses its energy is a random walk of $N=\lambda_{ee}/\lambda_{step}$, $\lambda_d = \sqrt{N} \lambda_{step}=\sqrt{\lambda_{step}\lambda_{ee}}$\cite{luciani1983,daviesNonlocalSuppressionBiermann2023}.
	
	Luciani \emph{et al.} took the total directionality step as $\lambda_{step}=\lei Z/(Z+1)$, which assumes that for $Z=1$ the electrons halve the length \cite{luciani1983}. This gives $\lambda_d = \sqrt{\lambda_{step}\lambda_{ee}}=\lei Z/\sqrt{Z+1}$.
	
	Epperlein \& Short assumed $\lambda_d = \sqrt{\lambda_{\kappa} \lambda_{ee}} = \lei \sqrt{Z(Z+0.24)/(Z+4.2)}$ (using their values for $\lambda_{\kappa}$). Walsh \& Sherlock adopted the same form for kinetic suppression of heat-flux\cite{walshKineticCorrectionsHeatflow2024}.
	
	This work measures the delocalization length relevant for heat-flow from the kinetic response computed with the full electron-electron collision operator:
	\begin{equation}
		\lambda_d = \lei \frac{Z}{\sqrt{Z+4.45}}
		\label{eq:lambdad}
	\end{equation}
	Inverting the random walk gives $\lambda_{step}=\lambda_d^2/\lambda_{ee}$, which will prove useful for understanding how magnetization relocalizes the plasma:
	\begin{equation}
		\lambda_{step} = \lambda_{ei} \frac{Z}{Z+4.45} \label{eq:lam_step}
	\end{equation}
	The form of $\lambda_{step}$ has a simple interpretation in the two-carrier picture of the previous paper\cite{walsh_paper0}, where heat is carried by fast electrons streaming down the gradient and slower electrons return to cancel the particle flux. e-e collisions drag on the fast heat carriers, adding to their e-i collision rate. The e-i rate scales with $Z$ while the e-e rate does not; adding the two rates gives $\lambda_{step}/\lambda_{ei}=Z/(Z+c)$. The constant $c=4.45$ itself is measured rather than predicted.

	\begin{figure}[tb]
		\includegraphics[width=\linewidth]{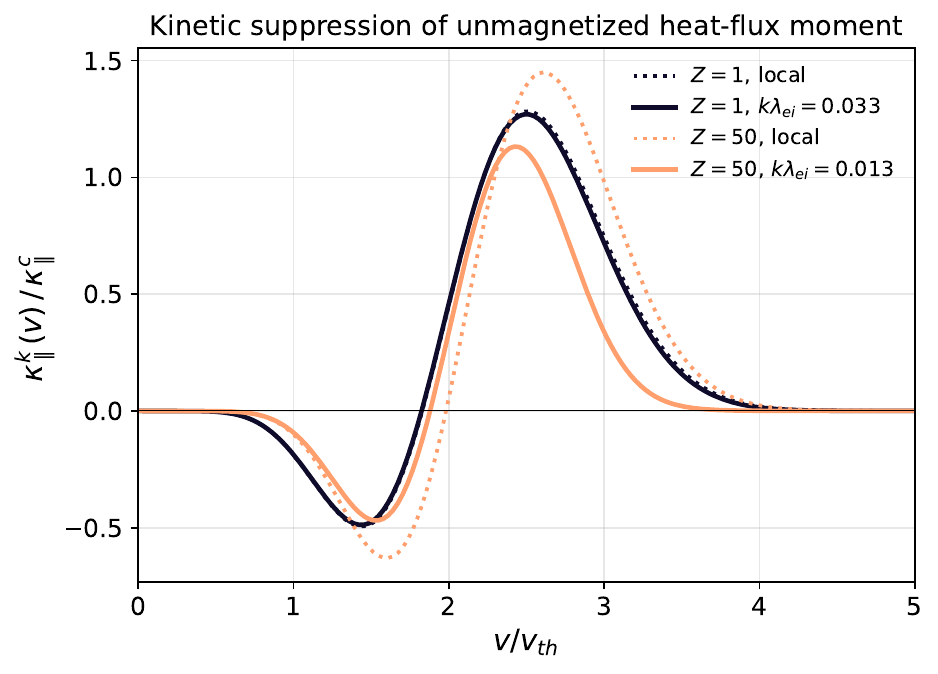}
		\caption{\label{fig:fig27_moment_suppression} Classical (dotted) and kinetic (solid) heat-flux moment against electron velocity for $Z=1$ (black) and $Z=50$ (orange). For $Z=1$ a non-locality level of $k\lei=0.033$ does not affect the thermal conductivity. For $Z=50$, $k\lei=0.013$ reduces the magnitude and shifts the carrier electrons to slower velocity, despite this seemingly being a lower level of non-locality. This motivates using a delocalization length $\lambda_d$ rather than $\lei$. }
	\end{figure}

	\begin{figure*}[tb]
		\includegraphics[width=\textwidth]{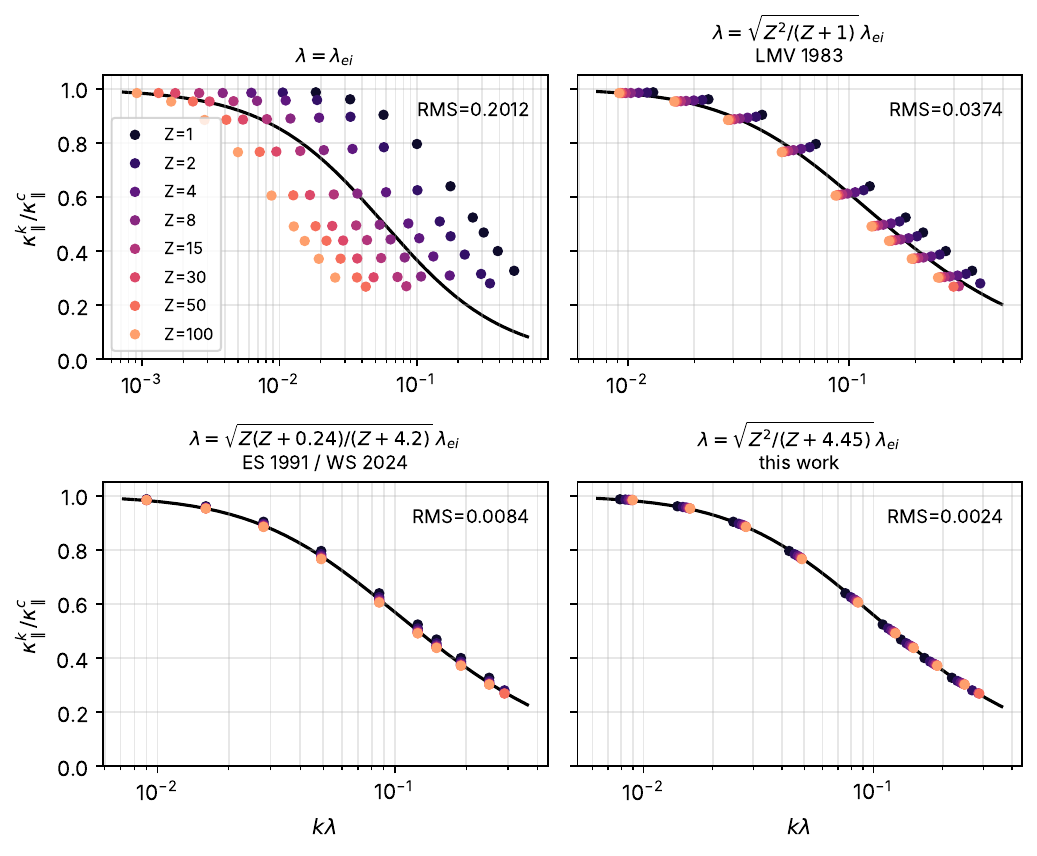}
		\caption{\label{fig:1_collapse} Unmagnetized heat-flow suppression ($\kappa_{\parallel}^k/\kappa_{\parallel}^c$) against non-locality metric ($k \lambda$) for each candidate non-locality length-scale $\lambda$. The black lines in panels (a)--(c) is fit to that panel's VFP data, with the RMS error given at the top; in (d) it is $\Gamma_q$ (equation \ref{eq:Gam}). Cases: a) electron-ion mean free path. b) LMV 1983\cite{luciani1983} $\lambda/\lambda_{ei} = \sqrt{Z^2/(Z+1)}$. c) ES 1991 \cite{epperleinPracticalNonlocalModel1991} $\lambda/\lambda_{ei} = \sqrt{Z(Z+0.24)/(Z+4.2)}$. d) this work $\lambda/\lambda_{ei} = \sqrt{Z^2/(Z+4.45)}$. }
	\end{figure*}
	\subsection{Magnetization length-scale \label{sec:mag_length}}
	An electron in a magnetic field no longer travels in straight lines across the field between collisions; it is confined to a Larmor radius $r_L = v_\perp/\omega_e$, where the subscript $\perp$ denotes velocity perpendicular to the magnetic field. The magnetic field does no work on the electron, but it does confine the mean free path between collisions. Writing this per electron velocity \cite{brantov2003}:
	\begin{equation}
		\lambda_{\perp}(v)  \;=\; \frac{\lambda(v)}{1 + \Big(\frac{\lambda(v)}{r_L(v)}\Big)^2}
		\label{eq:lambdadperp_v}
	\end{equation}
	
	With Braginskii's Hall parameter $\chi=\omega_e \tau_B$ and $\tau_B = 3\sqrt{\pi}\tau_{ei}/4$, the Hall parameter is $\chi=(3\sqrt{\pi}/4) \lei/r_L$; this is a comparison of unmagnetized e-i scattering length to the Larmor radius. However, it does not include the e-e collision contribution.
	
	The preceding work defined the carrier Hall parameter for thermal conductivity (equation \ref{eq:psi_kap})\cite{walsh_paper0}. Writing $\psi_{\kappa}=\lambda_s/r_L$ gives the classical scattering length of the heat-carrying electrons:
	\begin{equation}
		\lambda_{s} = 13.2\,\ \lei\, \frac{Z+0.17}{Z+5.88}  \label{eq:lam_s}
	\end{equation}
	where $13.2 = 9.94\times3\sqrt{\pi}/4$.
	
	This paper finds delocalization to be magnetized at a different rate to the classical conductivity. The magnetized delocalization length is measured in this paper:
	\begin{equation}
		\lambda_{d\perp}(Z,\psi_d)  \;=\; \frac{\lambda_{d}(Z)}{1 +\psi_d^{1.6}}
		\label{eq:lambdadperp}
	\end{equation}
	Where $\psi_d $ is the delocalization Hall parameter:
	\begin{equation}
		\psi_d \;=\; 13.4 \frac{Z}{Z+4.45} \chi \propto \lambda_{step} \chi \label{eq:psid}
	\end{equation}
	Note that the delocalization Hall parameter does not require any new fits to data; $\lambda_{step}$ was already found from the unmagnetized delocalization in section \ref{sec:unmag_length}. 
	
	The prefactor of $\psi_d$ ($13.4$)  is chosen for convenience; $\psi_d = 1$ is defined such that magnetization halves the delocalization length, which is the same convention as $\psi_{\kappa}=1$ halving $\kappa_{\bot}^c$. The prefactor also has a physical interpretation. In the Lorentz limit, $\psi_d =13.4\chi$ is the Hall parameter of the electrons at the peak of the heat-flux weight: $\chi(v)$ at $v=2.62v_{th}$ is $0.752\times2.62^3\chi=13.5\chi$ (equation \ref{eq:chi_v}). Note that this uses the peak of the weight rather than the carrier-averaged velocity ($2.48v_{th}$). $\psi_d$ therefore measures the magnetization of the electrons that carry the most heat.
	
	$\lambda_{d\perp}$ has a magnetization dependence of $\psi_d^{1.6}$, which is lower than the exponent $2$ predicted by the per-velocity form in equation \ref{eq:lambdadperp_v}. This is due to the feedback of magnetization itself lowering the carrier velocity (as shown in figure \ref{fig:carrier_map}) thereby making the plasma harder to further magnetize. This is discussed in more detail in previous work explaining the $\psi_{\kappa}^{1.5}$ dependence of $\kappa_{\bot}^c$\cite{walsh_paper0}.
	
	The Righi-Leduc heat-flux is found to delocalize faster than the perpendicular heat-flux. This can be understood most simply by comparing carrier velocities in figure \ref{fig:carrier_map}; Righi-Leduc is carried by higher energy electrons.
	
	However, this under-sells the inherent connection between perpendicular heat-flux and Righi-Leduc; Righi-Leduc suppresses like the perpendicular heat-flux to the second power: $\kappa_{\wedge}^k/\kappa_{\wedge}^c \approx (\kappa_{\bot}^k/\kappa_{\bot}^c)^2$.  The Righi-Leduc heat-flux originates from rotated perpendicular heat-flux (with the approximate relation $\kappa_{\wedge}^c\approx \psi_\kappa \kappa_{\bot}^c$ from the simple carrier models of coefficients\cite{walsh_paper0}); it inherits the perpendicular suppression, plus another factor of suppression from the carrier Hall parameter $\psi_\kappa$ being reduced.

	\section{VFP Simulations \label{sec:VFP}}
	
	The	Vlasov--Fokker--Planck code K2\cite{sherlockComparisonNonlocalElectron2017,sherlockSuppressionBiermannBattery2020,shaffer2023} is used to generate data at varying levels of non-locality. The setup used here is one dimensional with frozen ions and a fixed uniform magnetic field strength.
	
	The electron velocity distribution is expanded in spherical harmonics and truncated at $f_6$, leaving the $f_0$, $\mathbf{f}_1$, $f_2$, $\mathbf{f}_3$, $f_4$, $\mathbf{f}_5$ and $f_6$ system of equations. This expansion separates the electron speeds from their direction.  The most important harmonics for this work are $f_0$ and $\mathbf{f}_1$.
	
	$f_0(v)$ is the isotropic electron distribution: how many electrons have a speed $v$ regardless of their direction, with $f_0$ being centered at the ion rest frame. $f_0$ carries information about the electron density ($4\pi \int v^2 f_0 dv$) and temperature/energy ($\sim\int v^4 f_0 dv$). $\mathbf{f}_1(v)$ is the directional imbalance: if more electrons at a given $v$ are traveling in $+x$ than $-x$ then $f_{1x}(v)$ is positive. $\mathbf{f}_1(v)$ is a vector holding directional information such as current ($\sim \int v^3 f_{1x} dv$) and heat-flow ($\sim \int v^5 f_{1x} dv$).
	
	Neither $f_2$ nor higher harmonics contributes to the density, temperature, current or
	heat-flow; they enter only indirectly through their feedback on $\mathbf{f}_1$. Because
	$\mathbf{f}_2$ is itself of order $k v / \nu_{ei}(v)$ relative to $\mathbf{f}_1$, that
	feedback is second order, and the classical transport coefficients are
	recovered exactly at any truncation. The effect of truncation is therefore purely
	non-local, vanishing in the local limit and growing with non-locality. Appendix \ref{sec:regime_trunc} demonstrates the convergence of expansion up to $f_6$ and the importance of $f_2$ and $f_3$ to the suppression factors calculated in this paper. 
	
	Further discussion of the validated range of this paper can be found in appendix \ref{sec:regime}. Linearity is shown to hold through demonstration of superposition in appendix \ref{sec:regime_superposition}.

	\subsection{Collisions \label{sec:VFP_coll}}
	The ions are assumed to be stationary and infinitely massive relative to the electrons, such that electron--ion collisions randomize the electron direction but leaves the speed unchanged. Therefore, e-i collisions do not affect $f_0$. Instead e-i collisions act to damp $\mathbf{f}_1$, the directional preference of the electrons. This happens at a rate $\nu_{ei} \sim Zn_e/v^3$ and is trivial to include in VFP simulations, since the ions affect each electron velocity independent of the other electron velocities. With no e-e collisions the mean free path $\lambda_{ei} = v/\nu_{ei}$ then scales as $v^4$.
	
	The electron--electron collision operator is more complex, for two reasons. First, it acts on both $f_0$ and $\mathbf{f}_1$. Second, it results in electrons of one velocity depending on electrons at a different velocity (which shows up computationally as derivatives in velocity space). This complexity is often avoided in VFP codes by using a rescaled e-i collision frequency presented by Epperlein \& Short $\nu_{eff} = \nu_{ei}(Z+4.2)/(Z+0.24)$\cite{epperleinPracticalNonlocalModel1991,daviesNonlocalSuppressionBiermann2023,brodrickTestingNonlocalModels2017,kingham2004,sherlockSuppressionBiermannBattery2020}. While this rescaling can reproduce the classical thermal conductivity\cite{epperleinPracticalNonlocalModel1991} it does not reproduce the electron distribution of a more detailed collision operator. While the Epperlein \& Short approximation at $Z=1$ gives an effective collision frequency of $4.19\nu_{ei}$ across all electron velocities, the full collision operator in local conditions gives $\nu_{eff}(v/v_{th} = 1.25)\approx1$ and $\nu_{eff}(v/v_{th} = 3.5)\approx4.19$, rising with electron velocity. 	Appendix C of prior work\cite{walsh_paper0} demonstrates in detail how an effective collision frequency does not reproduce the change in electron carrier velocity required for evaluating non-locality.
	
	This paper retains the full electron--electron collision operator \cite{tzoufrasVlasovFokkerPlanck2011}, which acts on all harmonics. $f_0(v)$ is relaxed to Maxwellian through two e-e collision terms: velocity-space diffusion and drag from slower electrons. The e-e collisions on $f_0(v)$ move the energy between the electron population.
	
	e-e collisions acting on $\mathbf{f}_1$ governs the exchange of momentum in the population. Two types of terms are present: the 'test particle' terms reacting to the background ($f_0$ acting on $\mathbf{f}_1$), and the 'field particle' terms ($\mathbf{f}_1$ acting on $\mathbf{f}_1$). The test particle terms include pitch-angle scattering as well as drag and diffusion on the anisotropy; alone they do not conserve electron momentum. The field particle terms then restore this parity by including the reaction of the background to the test particle.
	
	The
	complete operator used here conserves momentum to $3\times10^{-5}$ of the
	electron--ion drag and recovers $\beta_\parallel$ to within $1.6\%$ of
	Epperlein \& Haines\cite{epperleinPlasmaTransportCoefficients1986} and
	$0.2\%$ of Sadler-Walsh-Li\cite{sadlerSymmetricSetTransport2021}, with
	$\kappa$ recovered to within $4\%$ across $Z=1$--$100$, comparable to the
	quoted accuracy of the fits themselves.

	\begin{figure}[tb]
		\includegraphics[width=\linewidth]{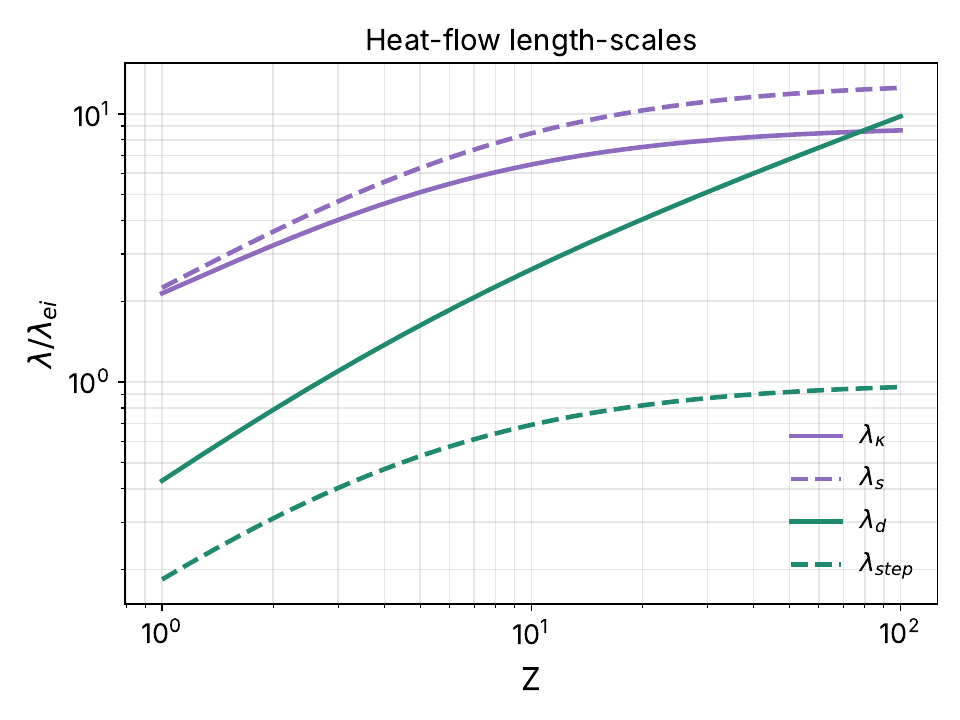}
		\caption{\label{fig:fig25_z_dependence} Impact of e-e collisions on different length-scales from section \ref{sec:theory}. Purple denotes classical length-scales, green for kinetic.  $\lambda_\kappa$ is the thermal conductivity scattering length, which is least affected by ionization $Z$. The ratio of $\lambda_s$ to the Larmor radius $r_L$ sets the carrier Hall parameter for thermal conductivity. The delocalization length $\lambda_d$ is the net distance an electron travels on a random walk between collisions before losing its energy, which can be compared with $L_T$ to assess non-locality. $\lambda_{step}=\lambda_d^2/\lambda_{ee}$ is the length of each step in that random walk; it tends to $\lei$ in the Lorentz limit and sets the delocalization Hall parameter, $\psi_d = 17.8\,\lambda_{step}/r_L$}
	\end{figure}
	
	\begin{figure}[tb]
		\includegraphics[width=\linewidth]{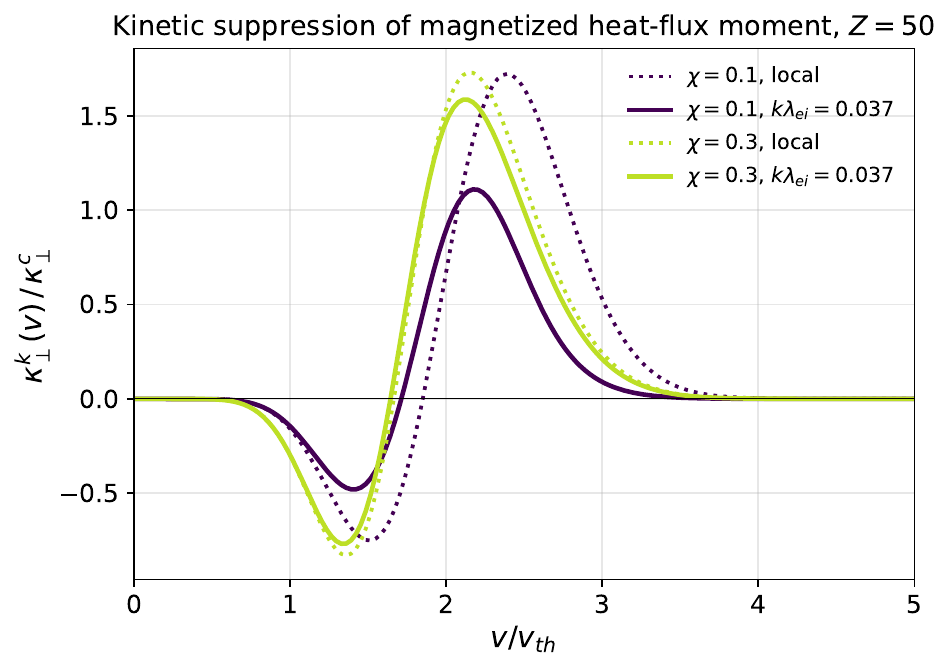}
		\caption{\label{fig:fig28_moment_suppression_mag} Classical (dotted) and kinetic (solid) heat-flux moment for $Z=50$ against electron velocity for $\chi=0.1$ (dark purple) and $\chi=0.3$ (lime green). Both kinetic simulations used $k\lei = 0.037$. The heat-carrying electrons at $\chi=0.1$ are heavily affected by kinetics, lowering the magnitude and lowering the carrier velocity. $\chi=0.3$ is much more robust to non-local effects. This motivates extending the delocalization length $\lambda_d$ to the magnetized regime.}
	\end{figure}
	
	\subsection{VFP Database}
	Each run is initialized with a temperature perturbation of the form $T=T_0(1 + \delta cos(kx)$) at uniform density. $k=2\pi/L_x$ is the perturbation wavenumber, where $L_x$ is the distance between perturbation peaks. This follows the standard
	Epperlein \& Short test problem\cite{epperleinPracticalNonlocalModel1991}. The nominal temperature perturbation used was $\delta = 0.01$, although runs were done spanning $\delta = 0.0025,0.08$ (see figure \ref{fig:3_amplitude}). The
	perturbation amplitude is chosen small enough that the response is linear; this is verified in section \ref{sec:regime_superposition} as well as figure \ref{fig:3_amplitude}.

	Transport coefficients $\underline{\underline{\kappa}}^k$ are extracted instantaneously at the location of maximum $|\nabla T_e|$. The kinetic suppression factors are then formed by dividing each measured
	coefficient by the local coefficient $\underline{\underline{\kappa}}^c$ evaluated at the same $Z$ and $\chi$. The local coefficients are re-calculated here to avoid interpolation errors on the published fits\cite{epperleinPlasmaTransportCoefficients1986,sadlerSymmetricSetTransport2021}; they are found to agree within the stated 4\% accuracy. The temperature perturbation is allowed to decay over time, with the thermal conductivity evaluated in time-dependent simulations once $\kappa^k/\kappa^c$ reaches a steady value. An alternative method would be to apply a heating operator to maintain the temperature perturbation over time\cite{daviesNonlocalSuppressionBiermann2023}; the difference between approaches is presented in appendix \ref{app:mode_definition} and shown to be small.
	
	The distribution function is discretized on a uniform grid in $x$ and $v$, with $N_x = 64$ spatial cells
	and $N_v = 256$ velocity cells spanning $0 \le v \le 8v_{th}$.
	Velocity-space resolution requirements are set by the carrier velocities: since transport is dominated by
	electrons at several times $\vth$, the grid must resolve the distribution
	well beyond the thermal peak, and $v_{max}$ must be chosen large enough
	that the transport integrals are converged rather than
	truncated. Grid convergence was tested, with $N_x = 128$ shifting the kinetic suppression of thermal conductivity by less than 2\% and $N_v=512$ by under 0.6\%, both below the 3\% accuracy of the classical transport coefficients\cite{epperleinPlasmaTransportCoefficients1986}.
	
	While the 1D VFP simulations are taken as the 'absolute truth' for the plasma transport response to non-locality, these are not the data points used to fit the constants in this paper. Instead, a quasi-static linear mode solver was validated against the full VFP simulations, and then used to generate a dense database of heat-flux suppression across $Z$, $\chi$ and $k \lambda_{d}$. This allowed for the 299 time-dependent 1D VFP datapoints to be extended to 118,041 quasi-static datapoints at higher resolution ($N_v = 512$), which improves the accuracy of the fits provided. The fits are compared against the 1D VFP simulations in all the figures of this paper, showing excellent agreement. More explanation of the linear mode solver can be found in appendix \ref{sec:eigensolver}.
	
	\section{Unmagnetized Kinetic Heat-flow \label{sec:unmag}}

	Section \ref{sec:carrier} demonstrated that plasma ionization $Z$ changes the susceptibility of heat-flow to non-locality through e-e collisions; the heat-carrying electrons are shifted to lower $v/v_{th}$ in low-$Z$ plasmas. However, the simulations used in that demonstration were classical in nature.
	
	Figure \ref{fig:fig27_moment_suppression} shows the unmagnetized heat-flux moment $\kappa_{\parallel}^k(v)$ normalized to the classical conductivity $\kappa_{\parallel}^c$ for 2 values of $Z$. The local equivalents are shown as dotted lines. The $Z=50$ plasma uses $k\lei=0.013$, where $k=2\pi/L_x$ is the temperature perturbation wavenumber (large $k$ means short perturbation wavelength). The high energy electrons carrying the flux are suppressed, which is often called tail depletion. The magnitude is lowered, such that $\kappa_{\parallel}^k/\kappa_{\parallel}^c = 0.60$. The carrier velocity is also lowered (seen by the shift to the left).

	Figure \ref{fig:fig27_moment_suppression} also plots $Z=1$ for a seemingly larger level of non-locality $k\lei=0.033$. However, in this case the anisotropic electron distribution is hardly affected by the non-locality with $\kappa_{\parallel}^k/\kappa_{\parallel}^c = 0.96$. The discrepancy demonstrates why using $\lei$ is not sufficient for parameterizing non-locality. A delocalization length that represents all ionizations is sought.

	
	Three length-scales from the literature are now tested against a suite of unmagnetized VFP data: 75 data points spanning 8 values of $Z$ and a range of $k$ (through changing the temperature perturbation wavelength). Each point is paired against a classical reference point in order to calculate the heat-flow suppression $\Gamma_q = \kappa_{\parallel}^k/\kappa_{\parallel}^c$, which spans $0.268$-$0.987$.  Note that the perturbation amplitude $\delta$ in this scan is kept constant at 1\%; this does not affect the results while in the linear regime, as discussed in section \ref{sec:FL}.
	
	The first candidate length-scale of interest is the electron--ion mean free path $\lei$; this is the typical length used to assess non-locality. Figure \ref{fig:1_collapse}(a) plots the simulated thermal conductivity suppression $\Gamma_q$ against $k \lei$. The dependence on $Z$ is substantial, with one case at $Z=1$ being essentially unsuppressed, while the equivalent $k \lei$ case at $Z=100$ has $\Gamma_q < 0.4$. A Knudsen number constructed with just the e-i mean free path therefore over-predicts suppression for low-$Z$ plasmas while under-predicting suppression for high-$Z$ plasmas.
	
	Figure \ref{fig:1_collapse}(b) and (c) then use lengthscales from LMV 1983\cite{luciani1983} and Epperlein \& Short 1991\cite{epperleinPracticalNonlocalModel1991} respectively. The latter case is utilized in the work of Walsh \& Sherlock 2024\cite{walshKineticCorrectionsHeatflow2024}. The length scales progressively improve from figure \ref{fig:1_collapse}(a) to (b) to (c), but there remains a small systematic dependence on plasma ionization.
	
	Fitting the linear-mode database gives $\lambda_d=\lei Z/\sqrt{Z+4.45}$. This is 12\% shorter at $Z=1$ than the length given by Epperlein \& Short\cite{epperleinPracticalNonlocalModel1991}, which is consistent with the 13\% expectation from the lowering of the carrier velocity with a full e-e collision operator (see appendix C from prior work\cite{walsh_paper0}).
	
	The kinetic heat-flow suppression is then:
	\begin{equation}
		\Gamma_q (k\lambda_d) \;=\; \frac{1+b k \lambda_d}{1 + b k \lambda_d + a (k \lambda_d)^2}
		\label{eq:Gam}
	\end{equation}
	The constants $a = 294.5$ and $b = 27.29$ are fitted to the same database. $k=2\pi / L_x$ is the wavenumber of the temperature perturbation applied, with $L_x$ being the periodic box length.

	The held-out VFP simulations are reproduced to an RMS of 0.0024 in $\Gamma_q$ (figure \ref{fig:1_collapse}(d)).
	
	Figure \ref{fig:fig25_z_dependence} plots the length-scales from section \ref{sec:theory} against ionization. Of importance here are $\lambda_{\kappa}$ and $\lambda_d$. The former is the distance that a heat-carrying electron travels before losing its directionality; this sets the classical thermal conductivity. The delocalization length $\lambda_d$ is the net distance an electron travels before losing its energy; this can be compared with the temperature length-scale $L_T$ to assess the impact of kinetics on transport. All of the length-scales deviate greatly from the e-i mean free path $\lei$.
	
	\section{Magnetized Kinetic Heat-flow \label{sec:mag}}
	
	Here $\lambda_d$ is extended into the magnetized regime for heat-flow perpendicular to a magnetic field. This is motivated by figure \ref{fig:fig28_moment_suppression_mag}, which plots the perpendicular heat-flux moment normalized to its classical counterpart $\kappa_\bot^k(v)/\kappa_\bot^c$ for $Z=50$. The solid lines use $k\lei=0.037$ at $\chi=0.1$ (black) and $\chi=0.3$ (green), while the dotted lines are the classical equivalents. As both values of $\chi$ use $Z=50$, they have the same delocalization length ($k\lambda_d = 0.25$); the results in section \ref{sec:unmag} suggest that they should be affected the same by non-locality. Figure \ref{fig:fig28_moment_suppression_mag} clearly shows the $\chi=0.3$ distribution being more robust to kinetics, with $\kappa_{\bot}^k/\kappa_{\bot}^c = 0.86$ compared with $0.52$ for $\chi=0.1$. The delocalization length perpendicular to the magnetic field is not the same as that parallel to the field.
	
	The metric for how the heat-flux is relocalized by a magnetic field is called the delocalization Hall parameter $\psi_d$. $\psi_d$ is chosen to be consistent with the unmagnetized delocalization length, and not fit to the data:
	\begin{equation}
		\psi_d = 13.4 \, \chi \, \frac{Z}{Z+4.45} \propto\lambda_{step} \chi
	\end{equation}
	where the prefactor $13.4$ is chosen such that $\psi_d = 1$ corresponds to the delocalization length being 50\% of its unmagnetized value. See section \ref{sec:mag_length} for more discussion on how $\lambda_{step}$ relates to $\lambda_{d}$.
	
	To find how the magnetized delocalization length $\lambda_{d\bot}$ depends on the delocalization Hall parameter, the linear-mode database is extended to $\chi=0$--$1$. 224 VFP simulations across 4 values of $Z$ ($1$, $4$, $15$, $50$), 8 values of $\chi$ ($0.01$-$1.0$) and a range of $k\lambda_d$ are held out for validation. The magnetized delocalization length then becomes:
	\begin{equation}
		\lambda_{d\bot} = \frac{\lambda_d}{1 + \psi_d^{1.6}}
	\end{equation}
	
\begin{figure}[tb]
		\includegraphics[width=\linewidth]{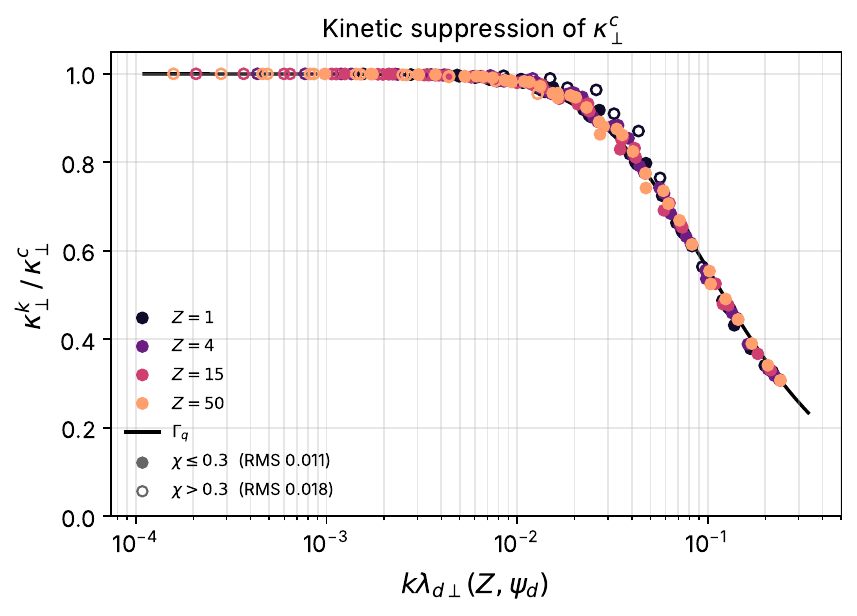}
		\caption{\label{fig:6_magnetised_collapse}  Kinetic suppression of perpendicular heat-flow across four ionizations and eight magnetizations ($\chi=0.01$-$1$), plotted against $k \lambda_{d\perp} (Z,\psi_d)$. The solid line is $\Gamma_q$ from equation \ref{eq:Gam}, with $\lambda_d$ and $\psi_d$ from table \ref{tab:lengths}. Filled symbols are $168$ simulations at $\chi \le 0.3$; open symbols are the $56$ runs at $\chi=0.5$ and $\chi=1$. All datapoints shown here were held out of the fits.}
	\end{figure}

	Figure \ref{fig:6_magnetised_collapse} shows the observed suppression of $\kappa_{\bot}^c$ for all 224 magnetized datapoints against $k\lambda_{d\bot}$. All $Z$ and $\psi_d$ dependence has been removed by the $\lambda_{d\bot}$ rescaling. The same suppression scaling from equation \ref{eq:Gam} is applied, using $\lambda_{d\bot}$ rather than $\lambda_d$:
	\begin{equation}
		\Gamma_{q\perp} = \Gamma_q(k\lambda_{d\perp})
	\end{equation}
	Despite using the same $\Gamma_q$ function as unmagnetized, the VFP simulations collapse to an RMS of 0.011 for $\chi\le0.3$ (filled symbols) and 0.017 for $\chi=0.5$ and 1 (open symbols), although many of the latter are essentially unaffected by kinetic suppression.
	
	It's worth dwelling on the difference between the carrier Hall parameter for thermal conductivity $\psi_{\kappa}$ and the delocalization Hall parameter $\psi_d$. The former comes from asking: 'are the classical electrons carrying  heat magnetized?'. The latter comes from posing the question: 'are the non-local electrons responsible for kinetic suppression magnetized?'  Since the 'kinetic electrons' are all of the fastest heat carriers, it is no surprise that they magnetize more easily than the bulk of the heat carriers.
	
		\subsection{Kinetic Righi-Leduc \label{sec:righi}}
		
		Figure \ref{fig:carrier_map} showed that Righi-Leduc is carried by higher velocity electrons than perpendicular heat-flow; this foreshadowed Righi-Leduc to be more strongly suppressed. This is borne out, with the kinetic Righi-Leduc coefficient measured through the $v^5 f_{1z}$ moment, i.e. the anisotropy of the electron distribution perpendicular to both the temperature gradient and the magnetic field. 
		
\begin{figure}[tb]
			\includegraphics[width=\linewidth]{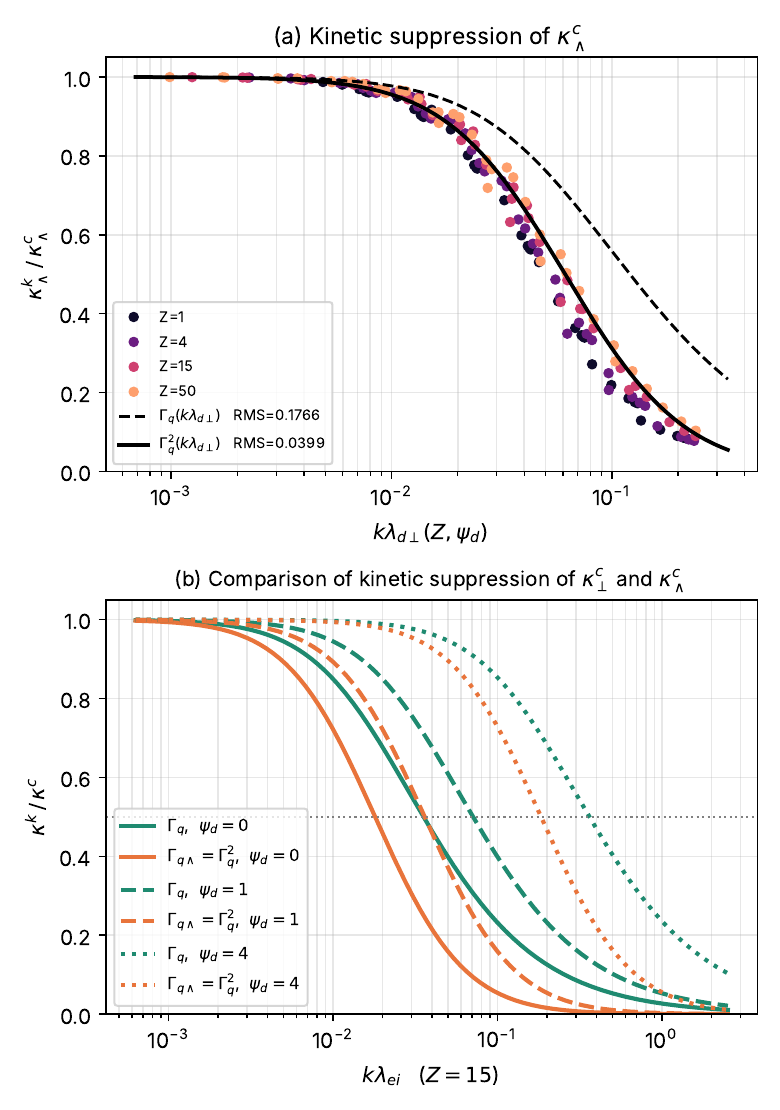}
			\caption{\label{fig:wedge_collapse}  (a): Kinetic Righi-Leduc coefficient normalized to its classical value across the magnetized VFP. The delocalization length developed for perpendicular heat-flux $\lambda_{d\bot}$ does a reasonable job at removing the $Z$-dependence. However, Righi-Leduc is more strongly suppressed than $\Gamma_{q\bot}$ suggests; $\Gamma_{q\bot}^2$ is a close fit and is physically-motivated.  (b): $\Gamma_{q}(k\lambda_{d\perp})$ and $\Gamma_{q\wedge}(k\lambda_{d\bot})$ against $k\lei$ for $Z=15$ and a range of magnetizations $\psi_d=0,1.0,4.0$. }
		\end{figure}

		Figure \ref{fig:wedge_collapse}(a) plots the kinetic Righi-Leduc coefficient normalized to its classical counterpart ($\kappa_{\wedge}^k/\kappa_{\wedge}^c$) for the magnetized VFP database; these are the same simulations as figure \ref{fig:6_magnetised_collapse} except analyzing the $v^5 f_{1z}$ moment rather than $v^5 f_{1x}$. Re-using the same delocalization length ($\lambda_{d\bot}$) and delocalization Hall parameter ($\psi_d$) as for the perpendicular heat-flux removes most of the $Z$-dependence of suppression. The suppression function for perpendicular heat-flux ($\Gamma_q(k\lambda_{d\bot})$) is plotted with a dashed line; it under-predicts Righi-Leduc suppression.
		
		Instead, Righi-Leduc suppression is found to follow:
		\begin{equation}
			\Gamma_{q\wedge}(k\lambda_{d\bot}) = \Gamma_{q}(k\lambda_{d\bot})^2
		\end{equation}
		The reasoning behind the power of $2$ is discussed in more detail in section \ref{sec:mag_length}. If the exponent is instead fit to the database, the RMS error in figure \ref{fig:wedge_collapse}(a) lowers from 0.04 to 0.02, with the exponent following $p = 1.87 (Z+10.8)/(Z+7.9)$. The power of 2 is retained: it follows from the single-carrier argument and needs no additional constants.
		
		Figure \ref{fig:wedge_collapse}(b) compares $\Gamma_q$ and $\Gamma_{q\wedge}$ across $k\lei$ at $Z=15$ for various magnetizations; Righi-Leduc is consistently more suppressed by kinetics than its perpendicular counterpart.

		\begin{figure}[tb]
			\includegraphics[width=\linewidth]{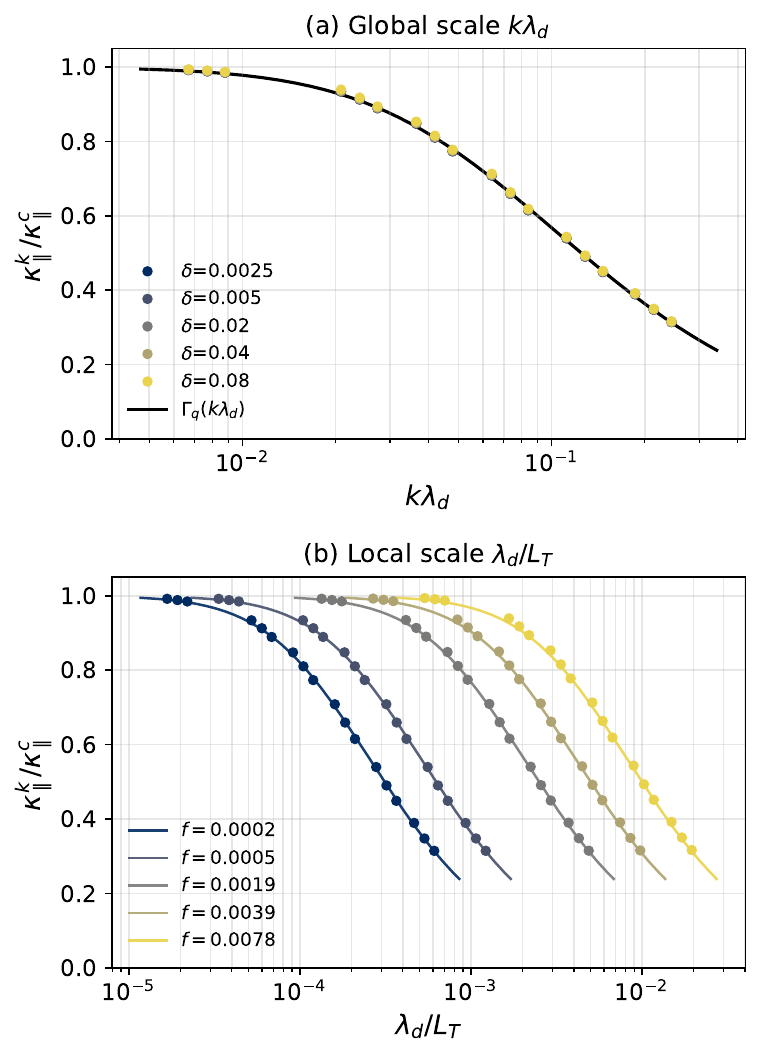}
			\caption{\label{fig:3_amplitude} Unmagnetized heat-flow suppression across a range of temperature perturbation amplitudes ($\delta$= 0.25\% to 8\%) viewed in 2 ways. (a) Uses the global scale ($k \lambda_d$), allowing collapse onto one form. However, this is not practical for implementation into a hydrodynamics code, where each cell only knows local conditions. (b) Uses the local scale ($\lambda_d / L_T$), which then requires a tuned flux limiter to fit the data points, with the flux limiter value scaling linearly with $\delta$.  }
		\end{figure}

		\section{Flux Limiters \label{sec:FL}}
		
		Since 1975 'flux limiters' have been a staple of plasma physics modeling\cite{malone1975}, restricting the electron heat-flow\cite{spitzerTransportPhenomenaCompletely1953} to a fraction of its free-streaming value. This is equivalent to placing the following limit on the thermal conductivity\cite{walshKineticCorrectionsHeatflow2024},
		\begin{equation}
			\kappa_\parallel ^c \le f_m \frac{L_T}{\lei} ,
			\label{eq:kap_lim}
		\end{equation}
		where $f_m$ is the 'min' flux limiter with typical values spanning $0.05-0.2$. It is widely agreed that the value of $f_m$ is not fixed and should be 'tuned' in an attempt to make radiation-hydrodynamics simulations better match laser-driven experiments \cite{jones2017,farmerHeatTransportModeling2018,doi:10.1063/5.0005776,farmer2021}.
		
		While the entire field would like to move beyond this crude approximation, VFP simulations are too computationally prohibitive for large-scale plasma physics experiments. Thermal flux limiters remain the norm. This section uses the findings of this paper to improve their use.
		
		\subsection{Global scale versus local scale \label{sec:FL_scale}}
		
		With excellent agreement between the suppression functions ($\Gamma_{q}=\kappa_{\parallel}^k/\kappa_{\parallel}^c$, $\Gamma_{q\bot}=\kappa_{\bot}^k/\kappa_{\bot}^c$, $\Gamma_{q\wedge}=\kappa_{\wedge}^k/\kappa_{\wedge}^c$) and the VFP data shown in figures (\ref{fig:1_collapse}(d), \ref{fig:6_magnetised_collapse}, \ref{fig:wedge_collapse}(a)) one might assume kinetic suppression of conductivity to be 'solved'. This is not the case.
		
		The results in this paper reproduce the linear kinetic response of heat-flux. The metric for non-locality is $k \lambda_d$, where $k=2\pi/L_x$ is the wavenumber of the perturbation and $L_x$ is the distance between sinusoidal temperature perturbation peaks. The as-built model therefore knows the perturbation \emph{global scale}. This is unknown in a typical hydrodynamics code.
		
		Thermal conduction is (most simply) calculated in a simulation cell by using $\nabla T_e$ at the cell faces; cells only know the temperature of their neighboring cells. The means of approximating kinetic suppression of heat-flow uses the temperature length-scale $L_T = T_e/|\nabla T_e|$, which neither distinguishes between modes nor represents the global scale of individual modes. For a single mode the smallest temperature length-scale (which is at the location of peak heat-flux) is $L_T = 1/k \delta$, where $\delta$ is the relative temperature perturbation amplitude.
		
		Heat-flux suppression is independent of $\delta$ in the linear response, as long as the global scale ($L_x=2\pi/k$) is known. Figure \ref{fig:3_amplitude}(a) shows the unmagnetized heat-flow suppression against $k \lambda_d$ for simulations with $\delta$ spanning $0.25\%$-$8\%$; the results lay on top of one another, which is consistent with linear response.
		
		If instead the global scale is unknown, the temperature length-scale at a particular location must be used. Figure \ref{fig:3_amplitude}(b) shows the same data plotted against $\lambda_d/L_T$; the estimated suppression is now dependent on the $T_e$ perturbation amplitude. Hydrodynamics codes use the tunable flux limiter value $f$ to move between these curves.

		\subsection{Flux limiter formulation \label{sec:FL_form}}
		
		The first consideration is the length-scale to be used in the flux limiter. The results of this paper are unambiguous: the delocalization lengths $\lambda_{d}$ from table \ref{tab:lengths}.
		
		Note that (to the author's knowledge) only the Gorgon extended-MHD code \cite{ciardiEvolutionMagneticTower2007,chittendenRecentAdvancesMagnetohydrodynamic2009,walshSelfGeneratedMagneticFields2017} currently uses $\lambda_d$ (rather than $\lambda_{ei}$) to suppress heat-flow, although even Gorgon uses the Epperlein \& Short\cite{epperleinPracticalNonlocalModel1991} form for $\lambda_d$\cite{walshKineticCorrectionsHeatflow2024}.
		
		The second open question is the form of the limiter in equation \ref{eq:kap_lim}; there it is written as a hard cap on the maximum thermal conductivity, as is the orthodoxy in hydrodynamics codes:
		\begin{equation}
			\kappa^k = \min(\kappa^c , f_{m} L_T/\lambda_d) \label{eq:fluxlim_cap}
		\end{equation}
		Others have suggested a harmonic mean as more representative of kinetic simulations \cite{malone1975,brodrickIncorporatingKineticEffects2018,walshKineticCorrectionsHeatflow2024}:
		\begin{equation}
			\kappa^k = \frac{\kappa^c}{1+ \frac{\lambda_d}{f_{h} L_T}} \label{eq:fluxlim_harm}
		\end{equation}
		
\begin{figure}[tb]
			\includegraphics[width=\linewidth]{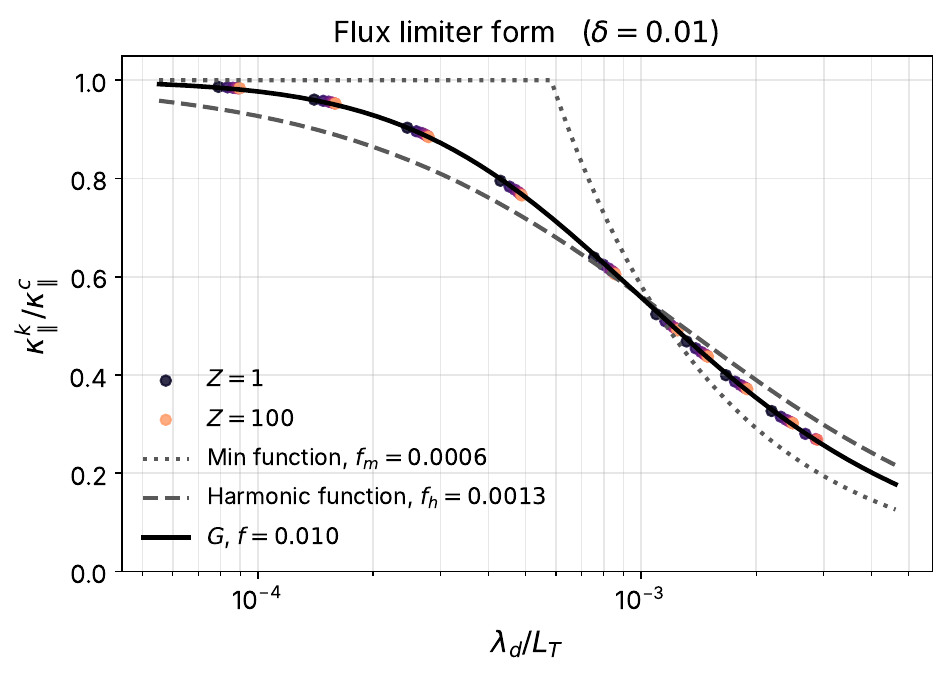}
			\caption{\label{fig:WS_compare}Comparison of different flux limiter forms against the unmagnetized VFP database. Dotted: minimum function flux limiter from equation \ref{eq:fluxlim_cap}. Dashed: harmonic flux limiter from equation \ref{eq:fluxlim_harm}\cite{walshKineticCorrectionsHeatflow2024}. Solid: flux limiter form proposed here in equation \ref{eq:fluxlim_G}, which is based off $\Gamma_q$.  }
		\end{figure}

		Figure \ref{fig:WS_compare} demonstrates the inadequacy of these forms by plotting them against the unmagnetized VFP database. The sharp features from the '$\min$' function are clearly not representing the smooth onset of kinetic suppression, as has been noted elsewhere\cite{walshKineticCorrectionsHeatflow2024}. The harmonic mean is an improvement, but overpredicts suppression at low $k\lambda_d$ and underpredicts at high $k\lambda_d$.

		
		This work instead proposes the following form for unmagnetized heat-flux, using equation \ref{eq:Gam}:
		\begin{equation}
			\kappa^k\;=\kappa^c\; G(f,\lambda=\lambda_d, L=L_T)
			\label{eq:fluxlim}
		\end{equation}
		\begin{equation}
			G= \frac{1+b \lambda/(f L)}{1 + b\lambda/(f L) + a (\lambda/(f L))^2}
			\label{eq:fluxlim_G}
		\end{equation}
		with a=294.5 and b=27.29. The delocalization length $\lambda_d$ can be found in table \ref{tab:lengths}.
		
		For magnetized heat-flow the conductivities become:
		\begin{equation}
			\kappa_{\parallel}^k\;=\kappa_{\parallel}^c\; G(f,\lambda=\lambda_{d}, L=L_{T})
			\label{eq:fluxlim_par}
		\end{equation}
		\begin{equation}
			\kappa_{\bot}^k\;=\kappa_{\bot}^c\; G(f,\lambda=\lambda_{d\bot}, L=L_{T})
			\label{eq:fluxlim_perp}
		\end{equation}
		\begin{equation}
			\kappa_{\wedge}^k\;=\kappa_{\wedge}^c\; G(f,\lambda=\lambda_{d\bot}, L=L_{T})^2
			\label{eq:fluxlim_wedge}
		\end{equation}
		with the constants unchanged from the unmagnetized suppression function: $a=294.5$ and $b=27.29$. 
		
		Note that all components use the same flux limiter value $f$, so there are no additional tuning parameters despite having 3 heat-flux components.
		
		A single scalar temperature length-scale $L_T = T_e/|\nabla T_e|$
		is built from the full temperature gradient at the cell face. This differs
		from the construction of Ref.~\cite{walshKineticCorrectionsHeatflow2024},
		which assigned each component its own length-scale from the corresponding
		gradient component ($L_{T\parallel}$, $L_{T\perp}$). That choice looks
		natural but fails a simple case: the
		unmagnetized result is not recovered as $B\to0$ unless $\nabla T_e$ is
		exactly parallel or perpendicular to $\mathbf{B}$. With one $L_T$ the
		limit is exact at every angle: as $B\to0$, $\lambda_{d\perp}\to\lambda_d$
		and $\kappa^c_\wedge\to0$, so equations
		\ref{eq:fluxlim_par}--\ref{eq:fluxlim_wedge} reduce to equation
		\ref{eq:fluxlim}. The VFP database validates this approach
		in the two geometries it contains, $\mathbf{B}\perp\nabla T_e$
		and $\mathbf{B}\parallel\nabla T_e$. Mixed geometries, where a
		gradient-weighted length-scale such as
		$L_T^{-2} = [\lambda_d^2(\nabla_\parallel T_e)^2 +
		\lambda_{d\perp}^2|\nabla_\perp T_e|^2]/(\lambda_d^2 T_e^2)$ would
		differ, are not tested here.
		
		The relation between the global-scale fit and its local-scale
		implementation is exact for a single mode: $\lambda_d/L_T = \delta\,k\lambda_d$
		at the point of steepest gradient, so equation \ref{eq:fluxlim} with $f=\delta$
		reproduces figure \ref{fig:3_amplitude}(b) with no additional fitting, and
		the flux limiter value scales linearly with the perturbation amplitude. For
		temperature structures of order-unity amplitude, $f\to1$ and the model has
		no free parameter. Two caveats keep $f$ tunable rather than
		constant. The linear response measured here is a $\delta\to0$ result, and
		its extrapolation to $\delta\sim1$ is untested; and a hydrodynamics code
		resolves a superposition of modes whose amplitudes it cannot separate. The
		value of $f$ therefore retains a physical meaning as the effective
		amplitude of the temperature structures being resolved; its tuned
		range in the literature, $0.05$--$0.2$, is
		consistent with $\delta$ of order unity.

		\section{Conclusion \label{sec:conc}}
		
		Electron-electron collisions reduce the impact of kinetics on plasma transport through lowering the average velocity of electrons carrying heat-flux.
		The delocalization length $\lambda_d=\lei Z/\sqrt{Z+4.45}$ has therefore been formulated to remove the ionization dependence of kinetic suppression on heat-flow.
		
		Magnetization also makes plasma transport across field lines more classical in nature; $\lambda_{d\perp}=\lambda_d/(1+\psi_d^{1.6})$ represents this behavior, where $\psi_{d}$ is the delocalization Hall parameter. $\psi_d=13.4\chi Z/(Z+4.45)$ is inferred from $\lambda_d$ rather than fitted again. $\psi_d > \psi_{\kappa}$, demonstrating that the electrons responsible for non-locality are more easily magnetized than the heat-flux carriers.
		
		Non-locality of plasma transport can be assessed using the improved Knudsen number:
		\begin{equation}
			Kn_{d} = 8.28\frac{\lambda_{d\perp}}{L_T}
		\end{equation}
		where the prefactor is chosen such that the heat-flow is 50\% suppressed at $Kn_d = 1.0$. $Kn_d=0.1$ is essentially unaffected by kinetics (3.1\%) and $Kn_d = 10$ is highly suppressed (92.7\%).
		
		Righi-Leduc heat-flux is found to simply follow the perpendicular component's suppression to the power of 2 ($\Gamma_{q\bot}^2$). This is consistent with the picture that it originates from the perpendicular flux (which is suppressed by one factor of $\Gamma_{q\bot}$), and is itself suppressed by the same factor.
		
		These findings allow for an improved flux limiter to be implemented into radiation-hydrodynamics and magneto-hydrodynamics codes. It may also allow for an improved SNB model\cite{schurtz2000}, e.g. through removing arbitrary length multipliers\cite{brodrickTestingNonlocalModels2017,sherlockComparisonNonlocalElectron2017} and allowing for relocalization of the electron population by magnetization. 
		
		The following two papers in the series apply similar methodology to the kinetic suppression of Nernst advection\cite{brodrickIncorporatingKineticEffects2018,walshKineticCorrectionsHeatflow2024} and Biermann battery generation\cite{sherlockSuppressionBiermannBattery2020,daviesNonlocalSuppressionBiermann2023}; suppression is found to follow the ordering set by the electron population responsible for carrying each process. The simulations used in those pieces of work remain in the regime of linear perturbations ($\delta T_e\ll T_e$) with marginal non-locality ($k_T \lambda_d <0.15$). Extending the database into 'deep non-locality', as well as non-\emph{linearity}, is the subject of future work.

		\begin{acknowledgments}
			The author would like to thank Mark Sherlock for encouraging use of the K2 code, and valuable discussions on kinetic theory. Deserved appreciation also goes to George Zimmerman for noticing numerical issues with the Walsh \& Sherlock flux limiter implementation discussed in section \ref{sec:FL_form}.
			
			This work was performed under the auspices of the U.S. Department of Energy by Lawrence Livermore National Laboratory under Contract DE-AC52-07NA27344.
			
			This document was prepared as an account of work sponsored by an agency of the United States government. Neither the United States government nor Lawrence Livermore National Security, LLC, nor any of their employees makes any warranty, expressed or implied, or assumes any legal liability or responsibility for the accuracy, completeness, or usefulness of any information, apparatus, product, or process disclosed, or represents that its use would not infringe privately owned rights. Reference herein to any specific commercial product, process, or service by trade name, trademark, manufacturer, or otherwise does not necessarily constitute or imply its endorsement, recommendation, or favoring by the United States government or Lawrence Livermore National Security, LLC. The views and opinions of authors expressed herein do not necessarily state or reflect those of the United States government or Lawrence Livermore National Security, LLC, and shall not be used for advertising or product endorsement purposes.
			
			The data that support the findings of this study are available from the corresponding author upon reasonable request.
		\end{acknowledgments}

		\appendix

		\section{Assumptions and validated range \label{sec:regime}}
		
		The results in this paper are the linear response of a kinetic plasma to a single sinusoidal temperature perturbation. The results apply to marginally non-local plasmas, with future work studying further corrections in the deeply non-local regime ($k\lambda_{d}>0.15$). This section starts by addressing the accuracy of the suppression factor $\Gamma_q$ (equation \ref{eq:Gam}) and relevant length-scales (table \ref{tab:lengths}).

		The parameter range simulated is $Z=1$ to $100$, magnetization $\chi = 0$ to $1$ ($\psi_{\kappa} = 0$ to $8.9$, $\psi_{d} = 0$ to $12.3$), non-locality $k \lambda_d = 0.0079$ to $0.29$ and perturbation amplitude $\delta_0 = 0.0025$ to $0.08$. Within that range the suppression of heat-flow follows equation \ref{eq:Gam} to an RMS of $0.0024$ in $\Gamma_q$ unmagnetized, $0.011$ magnetized perpendicular ($\chi\le0.3$), and $0.040$ for Righi-Leduc, against the held-out VFP simulations.
		
		The worst deviation (VFP minus fit) for unmagnetized heat-flow is $-0.007$ in $\Gamma_q$ at $Z=1$, $k\lambda_d=0.075$. For magnetized perpendicular ($\chi\le0.3$) that is $-0.034$ at $Z=50$, $\chi=0.2$, $k\lambda_{d\bot}=0.047$. For Righi-Leduc the worst deviation is $-0.141$ at $Z=4$, $\chi=0.3$, $k\lambda_{d\bot}=0.063$.

		For $k\lambda_d > 0.15$ there is evidence of more complicated behavior emerging, with magnetization becoming dependent on $k$; therefore the results here are intended as applying to marginally non-local transport. 
		
		Appendix \ref{sec:regime_trunc} quantifies the importance of including harmonics above $f_1$ in this work and demonstrates convergence at $f_6$. Appendix \ref{sec:regime_superposition} shows that linear superposition applies to the results of this paper.  
		
		\subsection{Truncation of the harmonic expansion \label{sec:regime_trunc}}
		
		The simulations and linear mode solver in this paper truncate the spherical-harmonic
		expansion at $f_6$. Table \ref{tab:trunc} shows the effect of the truncation, comparing
		the suppression factors at truncations $f_1$--$f_6$ against $f_8$ across the linear mode
		database ($Z=1$--$100$, $\chi=0$--$3$). Truncation error vanishes in the local limit
		and grows with non-locality; it is largest for Righi-Leduc, which is carried by the
		fastest electrons. $f_1$ truncation is not sufficient for these results, with errors in 
		$\Gamma_{q\perp}$ up to $0.06$ and $\Gamma_{q\wedge}$ up to $0.24$; at $f_6$ the
		truncation error is below $3\times10^{-4}$ throughout the fitted range.
		
		\begin{table}[tb]
			\caption{\label{tab:trunc} Maximum absolute error in the suppression factors
				when truncating at $f_L$, relative to $f_8$, over $k\lambda_d\le0.1$ and $\le0.28$.
				$\Gamma_{q\perp}$ includes the unmagnetized $\Gamma_q$.}
			\begin{ruledtabular}
				\begin{tabular}{ccccc}
					& \multicolumn{2}{c}{$k\lambda_d\le0.1$} & \multicolumn{2}{c}{$k\lambda_d\le0.28$} \\
					\cline{2-3} \cline{4-5}
					$L$ & $\Gamma_{q\perp}$ & $\Gamma_{q\wedge}$ & $\Gamma_{q\perp}$ & $\Gamma_{q\wedge}$ \\
					\colrule
					1 & $0.061$            & $0.22$             & $0.064$            & $0.24$             \\
					2 & $7.7\times10^{-3}$ & $0.049$            & $0.018$            & $0.095$            \\
					3 & $7.0\times10^{-4}$ & $5.9\times10^{-3}$ & $3.0\times10^{-3}$ & $0.032$            \\
					4 & $5.6\times10^{-5}$ & $3.9\times10^{-4}$ & $5.3\times10^{-4}$ & $8.5\times10^{-3}$ \\
					5 & $6.9\times10^{-6}$ & $9.7\times10^{-6}$ & $7.7\times10^{-5}$ & $1.7\times10^{-3}$ \\
					6 & $4.4\times10^{-6}$ & $1.6\times10^{-6}$ & $8.7\times10^{-6}$ & $2.8\times10^{-4}$ \\
				\end{tabular}
			\end{ruledtabular}
		\end{table}
		
		\subsection{Decaying and stationary temperature modes \label{app:mode_definition}}
		
\begin{figure}[tb]
			\includegraphics[width=\linewidth]{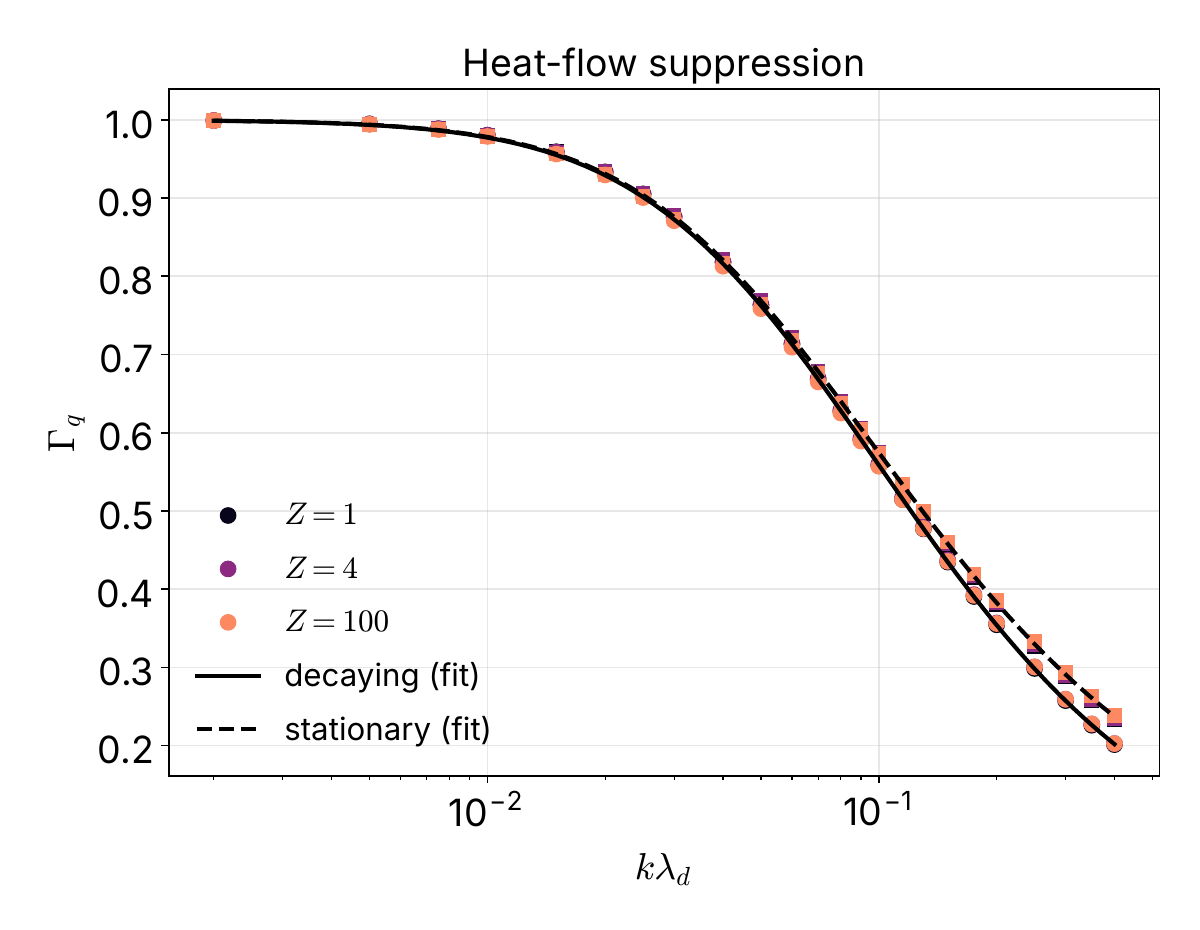}
			\caption{\label{fig:mode_definition} Heat-flow suppression for decaying (circles, solid line) and stationary (squares, dashed line) temperature modes, for $Z=1$, 4 and 100.}
		\end{figure}

		The temperature perturbations in this paper are decaying modes: the perturbation relaxes at the rate set by heat flow. An alternative is a stationary mode, where the perturbation is held fixed by heating\cite{daviesNonlocalSuppressionBiermann2023}. Figure \ref{fig:mode_definition} compares the heat-flow suppression for the two cases. They agree for $k\lambda_d < 0.03$, but the stationary mode is less suppressed at larger $k\lambda_d$; the difference is 3\% at $k\lambda_d=0.1$ and 18\% at $k\lambda_d=0.4$, with little dependence on $Z$. The stationary results are fit by:
		\begin{equation}
			\Gamma_q^{s} = \left[\frac{1 + 24.74\,k\lambda_d}{1 + 24.74\,k\lambda_d + 353.9\,(k\lambda_d)^2}\right]^{0.789}
		\end{equation}
		
		This difference bounds the range over which $\Gamma_q$ depends on the temperature gradient alone. For $k\lambda_d \lesssim 0.1$ the heat flow is within a few percent of equation \ref{eq:Gam}.
		
		\subsection{Superposition \label{sec:regime_superposition}}
		
		Until this section all simulations used a single-mode temperature perturbation. Realistic plasmas naturally contain a superposition of modes.
		
		The observed behavior of the single-mode simulations is asserted to be the 'linear response'; if this is true, multiple modes are expected to react as a superposition of independent modes with no mode coupling.
		
		Two-mode simulations were initialized using the following temperature profile:
		\begin{equation}
			T_e = T_0(1 + \delta_1 sin(k_1 x) + \delta_2 sin(k_2 x))
		\end{equation}
		These were run at $Z=1$ and $50$ using $k_1 \lambda_d = 0.028$, $0.086$ and $0.250$ and a second mode at twice and three times that wavenumber. The mode amplitudes were kept the same for each run ($\delta_1 = \delta_2$), which is deemed the strictest test of linear response. $\delta_1$ was varied between simulations at $\delta_1 = 0.005$ and $0.020$. These runs were unmagnetized.
		
		Conceptually, superposition requires that mode $n$ ($=1,2$ for two-mode) has its own thermal conductivity ($\kappa^k_1$) that acts exclusively on the temperature perturbation of that mode $q_n = -\kappa^k_n \partial (T_0 \delta_n sin(k_n x))/\partial x$. Each mode's conductivity is extracted from the Fourier components of the heat-flow and temperature, and compared with a single-mode run at identical conditions.
		
		The conductivities agree to $2.0\%$ (relative) on average at $\delta = 0.005$ and $2.4\%$ at $\delta = 0.020$. The decomposed modes acted as if the other modes were not present. Linear superposition holds.

		\section{Linear mode solver \label{sec:eigensolver}}
		
		The full VFP simulations (section \ref{sec:VFP}) evolve a temperature perturbation in time until it settles into a steady decay; then the heat-flux is read out of the simulation and compared to the classical value. The linear mode solver instead finds the state directly, as the slowest-decaying temperature eigenmode of the linearized VFP equations at a given $Z$, $\chi$ and $k$. 
		
		The linear mode solver is built from the same collision operators, velocity
		grid and harmonic truncation as the VFP code, with the harmonics $\mathbf{f}_1$--$f_6$ treated
		as quasi-static. As in the
		VFP database, each mode is normalized by a nearly local mode ($k\lambda_d=10^{-4}$)
		at the same $Z$ and $\chi$.
		
		Evaluated at the same conditions as each of the 299 VFP simulations,
		it reproduces the suppression factors to a median of $1\times10^{-4}$ for $\Gamma_q$
		(maximum $0.002$), and $8\times10^{-4}$ for both $\Gamma_{q\perp}$ (maximum $0.003$)
		and $\Gamma_{q\wedge}$ (maximum $0.004$), an order of magnitude below the accuracy of
		the fits. 
		
		Each calculation takes seconds, which allows for the generation of $>10^5$ datapoints rather than several hundred using the time-evolved VFP calculation. The database spans $Z=1$--$100$, $\chi=0$--$3$ and $k\lambda_d=0.002$--$0.5$
		(logarithmically spaced). Higher velocity resolution ($N_v=512$) is used for the fits, while a separate database at moderate resolution ($N_v=256$) is used for comparison with the full VFP data. The fits use $k\lambda_d\le0.3$, with the magnetized
		constants fitted on $\chi\le0.3$ only. The author hopes these will serve as reference fits for the linear kinetic response of plasma transport.

		\section{Kinetic carrier velocities \label{sec:kinetic_carrier}}
		
		Previous work\cite{walsh_paper0} fitted the classical electron carrier velocity for heat-flux $\langle v\rangle_{\rm c}$ as a function of $Z$ and $\psi_{\kappa}$; this was the function used to generate figure \ref{fig:carrier_map}.
		
		The carrier velocity is now extended to the kinetic regime using the kinetic database in this paper:
		\begin{equation}
			\frac{\langle v\rangle(Z,\chi,k)}{\langle v\rangle_{\rm c}(Z,\chi)} \;=\;
			\,0.775 + \frac{0.225}{1+\left(8.0\,k\lambda_{d\perp}\right)^{1.50}}
			\label{eq:v_mag_nonlocal}
		\end{equation}
		Where $	\langle v\rangle$ is the mean speed of $|v^5 f_{1x}|$ in units of $v_{th} = \sqrt{2T_e/m_e}$ and the subscript $c$ denotes the classical carrier velocity. The argument $k \lambda_{d\perp}$ (from equation \ref{eq:lambdadperp}) reduces to $k\lambda_{d}$ (from equation \ref{eq:lambdad}) in the unmagnetized case. Non-locality reduces the carrier velocity alongside classical magnetization and e-e collisions.
		
		The Righi-Leduc carriers, defined as the mean speed of $|v^5 f_{1z}|$, follow the same form. Classically they are faster than the perpendicular heat-flux carriers and are slowed less by magnetization\cite{walsh_paper0}. Under non-locality, normalizing each run by its classical reference,
		\begin{equation}
			\frac{\langle v\rangle_\wedge}{\langle v\rangle_{\wedge,c}} = 0.718 + \frac{0.282}{1+\left(8.8\,k\lambda_{d\perp}\right)^{1.50}},
			\label{eq:v_wedge_nonlocal}
		\end{equation}
		to an RMS of $0.005$. The Righi-Leduc carriers lose more of their speed than the perpendicular carriers ($0.718$ against $0.775$ in the non-local limit), consistent with Righi-Leduc being carried by faster electrons and so being more strongly suppressed (section \ref{sec:righi}).

		\bibliographystyle{aipnum4-2}
		\bibliography{library}

	\end{document}